\documentclass[12pt]{article}

\usepackage{latexsym}
\usepackage{mathrsfs}
\usepackage{amssymb}
\usepackage{eucal}

\newcommand{\euB}{\mathfrak{B}}
\newcommand{\euK}{\mathfrak{K}}
\newcommand{\euA}{\mathfrak{A}}
\newcommand{\euM}{\mathfrak{M}}

\begin{document}

\begin{titlepage}
\vspace{0.5truecm}

\begin{center}

{\Large \bf Quantum Logic and Non-Commutative Geometry}

\vspace{2.5cm}

P.A. Marchetti

\smallskip

{ \it Dipartimento di Fisica, Universit\`a degli Studi di Padova,

\smallskip

and

\smallskip

Istituto Nazionale di Fisica Nucleare, Sezione di Padova,

Via F. Marzolo, 8, 35131 Padova, Italia

\smallskip
e-mail: marchetti@pd.infn.it}

\vspace{1cm} R. Rubele

\smallskip
{ \it Dipartimento di Fisica ``E.R. Caianiello", Universit\`a
degli Studi di Salerno,

\smallskip

and

\smallskip

Istituto Nazionale di Fisica della Materia, Unit\`a di Salerno,

Via S. Allende, 84081 Baronissi (SA), Italia

\smallskip
e-mail: rubele@sa.infn.it }

\vspace{4.5cm}

\normalfont

\begin{abstract}

\vspace{0.5cm} \noindent We propose a general scheme for the
``logic'' of elementary propositions of physical systems,
encompassing both classical and quantum cases, in the framework
given by Non Commutative Geometry. It involves Baire*-algebras,
the non-commutative  version of measurable functions, arising as
envelope of the $C$*-algebras identifying the topology of the
(non-commutative) phase space. We outline some consequences of
this proposal in different physical systems. This approach in
particular avoids some problematic features appearing in the
definition of the state of ``initial conditions'' in the standard
($W^*$-)algebraic approach to classical systems.

\end{abstract}

\end{center}
 \vskip 1.0truecm

\end{titlepage}

\newpage

\baselineskip 6 mm
\section{Introduction}

\vskip 0.2truecm

\noindent In many respects Non-Commutative Geometry (NCG) \cite
{Connes1} appears as the most complete mathematical setting for a
unified description of quantum and classical physical systems,
besides being a source of some highly imaginative ideas in the
attempt of constructing a unified theory of fundamental forces
including gravity (see e.g.\cite{doplicher, connes2, froehlich,
witten} and references therein).

In this paper we propose a characterisation of the lattice of elementary
propositions, i.e. the ``logic'', of quantum and classical systems
which appears to fit naturally in the framework of NCG and solves
some problematic feature of the more standard $W$*-algebraic
approach (see e.g. \cite{haag, primas, redei, thirring}). In order
to keep the paper reasonably self-contained some basic notions
concerning $C$*- and $W$*-algebras used throughout the text are
given in the Appendix.

A root of Non-Commutative Geometry is the idea that one can
generalise many branches of standard functional analysis, such as
measure theory, topology and differential geometry, by replacing
the commutative algebras of functions over some space $X$ by a
suitable non-commutative algebra which may in a sense be
interpreted as the ``algebra of functions over a non-commutative
space".

In the commutative case one can consider various degrees of
regularity of the functions ranging from measurable, to
continuous, to smooth. The non-commutative analogue of the algebra
of complex bounded continuous functions is a $C$*-algebra, whereas
spaces of complex essentially bounded measurable functions
($\mathscr{L}^\infty$) are generalized by von Neumann algebras or,
in abstract form, $W$*-algebras. Algebraic generalizations of
spaces of smooth functions are pre-$C$*-algebras, i.e.
$\ast$-subalgebras of a $C$*-algebra closed under the holomorphic
functional calculus.
Probability measures on spaces of continuous functions find a non-commutative generalization in the concept of algebraic states, henceforth simply states: the linear positive normalized functionals on a $C^*$-algebra; in particular Dirac measures with support on one point are generalized by pure states, i.e. states that cannot be written as convex combinations of other states. (Notice that since a $C^*$-algebra is a Banach space, states are elements of its dual as they are continuous being bounded.)

A link with quantum theory appears when quantum mechanics is
interpreted as a ``mechanics over a non-commutative phase space"
in the spirit of Heisenberg and Dirac. If we consider a quantum
non-relativistic elementary particle with classical analogue i.e.
without internal degrees of freedom, the appropriate algebra of
``smooth functions, or observables, in phase space" is the Weyl
algebra generated by the bounded version,

\begin{equation}
e^{i \vec\alpha\cdot \vec q} e^{i \vec\beta \cdot \vec p} =
e^{i\vec\beta \cdot \vec p} e^{i \vec\alpha\cdot \vec q} e^{i
\hbar \vec\alpha\cdot \vec \beta \over 2},
\end{equation}

\noindent of the celebrated Heisenberg commutation relations:

\begin{equation}
q_i p_j - p_j q_i = i \hbar \delta_{ij},
\end{equation}

\noindent where $\alpha_i, \beta_i \in \mathbb{R}$ and $\{ q_i
\}^3_{i=1}$ and $ \{p_i\}^3_{i=1}$ are the ``coordinates" of the
``non-commutative phase space" corresponding respectively to
canonical coordinates of the underlying commutative classical
configuration space and their conjugate momenta. It turns out that
the corresponding $C$*-algebra of ``continuous
bounded observables" is
isomorphic to the $C$*-algebra $\euK ({\mathscr H})$ of compact
operators on an infinite dimensional separable complex Hilbert
space, ${\mathscr H}$, and the $W$*-algebra of ``bounded
measurable observables" is isomorphic to the algebra $\euB
({\mathscr H})$ of all bounded operators on ${\mathscr H}$. (The qualification ``continuous bounded'' used above is meant
to evocate the analogy with the commutative case and is not
referred to norm continuity of operators on Hilbert spaces, which
of course is equivalent to boundedness.)

A relation with quantum logic then appears as follows. It has been
recognised in the seminal work of Birkhoff and von Neumann
\cite{vonneumann}, that the system of elementary physical
propositions corresponding to yes-no experiments of quantum
mechanics can be represented as the complete
orthomodular\footnote{In an orthocomplemented lattice
$\mathscr{L}$ with partial order $\leq$ the orhomodularity can be
expressed as: if $a,b \in \mathscr{L}$ and $a \leq b$, then $b= a
\vee (a^\bot \wedge b)$.} lattice of closed
subspaces of a separable Hilbert space ${\mathscr H}$. (Actually, orthomodularity was not
introduced in \cite{vonneumann}, but by Piron \cite{piron1}; for
an historical comment see \cite{redei}.) Such lattice can be characterized also algebraically in terms of the associated orthogonal projectors, $p$, in
${\mathscr H}$, with the well known definitions of orthocomplement
${}^\bot$, meet $\wedge$ and join $\vee$ operations: $p^\bot ={\bf
1} -p$, $p_1 \wedge p_2 =$lim$_{n \rightarrow \infty}(p_1 p_2)^n =
$lim$_{n \rightarrow \infty}(p_2 p_1)^n$, $p_1 \vee p_2 =(p_1^\bot
\wedge p_2^\bot)^\bot$ and partial ordering defined by $p_1 \leq
p_2$ iff $p_1=p_1 \wedge p_2$. In turn, the projectors are the
self-adjoint elements of the von Neumann algebra  $\euB ({\mathscr
H})$ satisfying $p^2=p$. The set of projectors of any $W$*-algebra
has the structure of a complete orthomodular lattice with lattice operations defined algebraically as above. Therefore it
has been proposed to identify as a model for the propositional
lattice of physical systems the lattice of projectors of a
$W$*-algebra.

A classical system in this setting is given in terms of a
commutative $W$*-algebra; the corresponding lattice of
propositions is therefore distributive, i.e. a Boolean algebra.
Hence the transition from the classical to the quantum level
corresponds to the elimination of the commutativity postulate, due
to the existence of the universal constant $\hbar$, which is
replaced by $0$ in classical mechanics. More precisely, for a
classical particle the $W$*-algebra generated by the commutation
relation (1) with $\hbar =0$ is taken to be $\mathscr{L}^\infty
(\Omega, \omega_L)$ where $\Omega$ is the phase space and
$\omega_L$ is the Lebesgue measure on $\Omega$ which coincides
with the Liouville measure given in terms of the symplectic form.

Although the $W$*-approach has the great virtue of describing
classical and quantum systems and the related logics in a unifying
canonical scheme, it reveals some drawbacks in the definition of
states at the classical level. In the algebraic approach the states describe the ``states of
knowledge" of the observable quantities and pure states correspond
to maximal knowledge. However in classical systems points in phase space are
of zero $\omega_L$-measure and hence ``invisible" to
$\mathscr{L}^\infty (\Omega, \omega_L)$. Therefore, as already
noticed by von Neumann, it is not naturally defined in this setting the most fundamental state of classical
mechanics corresponding to a single point in phase space selecting ``initial conditions'' of the system; see \cite{halvorson} for
a more refined and recent analysis of the problem. Although this
fact could be attributed to the practical impossibility of a
precise measurement, it is at least philosophically somewhat
unnatural. For a related problem e.g. Teller \cite{teller} argued
that \textit{``if we believe that systems possess exact values for
continuous quantities, classical theory contains the descriptive
resources for attributing such values to the system, whether or
not measurements are taken to be imprecise in some sense''.}

Instead, points in phase space can be taken as support of Dirac measures and these are naturally defined as states on
$C_0 (\Omega)$, the $C$*-algebra of bounded continuous functions
on $\Omega$ vanishing at infinity, generated by the commutation
relations (1) with $\hbar =0$. However $C_0 (\Omega)$ does not
contain non-trivial projectors, since these are characteristic functions which are not continuous. Analogously the $C$*-algebra
of compact operators on a separable Hilbert space, $\euK({\mathscr
H})$, generated by (1) with $\hbar \not =0$, does not contain a
lattice of projectors even $\sigma$-complete, i.e. stable under a
countable number of meet and join operations, and this is the weakest
reasonable completeness to require in a logic, excluding ``unsharp''
approaches, see e.g. \cite{dallachiara} (we use the word
``complete'' to denote stability under an arbitrary, even non
countable, number of meet and join operations). On the other hand the pure
states on $\euK({\mathscr
H})$ are exactly in correspondence with the rays of ${\mathscr
H}$, as required on physical grounds. In fact the dual of $\euK({\mathscr
H})$ is isomorphic to the space of trace-class operators on ${\mathscr
H}$, the condition of positivity and normalization then identifies the states as the ``statistical matrices''. The pure states correspond to one dimensional projectors hence to rays, but this correspondence does not hold
for the pure states on  $\euB
({\mathscr H})$, which include also unphysical ``improper states''. ( To save the physically required correspondence in this case one has to restrict to the normal states, i.e those which are completely additive.)

Hence in a NCG setting as a natural framework to embed an algebraic model of elementary propositions one is
naturally looking for a ``space'' in general larger then the
$C$*-algebra of ``continuous bounded observables" $\euA$, but
smaller than the $W$*-algebra of ``essentially bounded measurable
observables", and containing a $\sigma$-complete orthomodular
lattice of projectors. Furthermore one would like this space still
to be some ``closure" of the $C$*-algebra $\euA$, which in the NCG
approach identifies the topology of the non-commutative phase
space and is taken as the basic algebra, identifying the space of physical
states.

This ``space'' in fact exists, it is called Baire*-algebra and can
be described in the above terminology as the $C$*-algebra of
(Baire) measurable bounded functions or observables on a generally
``non-commutative" space, and it is generated by $\euA$, as a
suitable enveloping algebra. We denote it by $\mathscr{B}(\euA)$.

We propose to identify the lattice of projectors of
$\mathscr{B}(\euA)$, denoted by $\mathbb{P}(\mathscr{B}(\euA))$,
as a model for the lattice of elementary propositions of the physical systems
described by $\euA$, and to identify the logical states $\phi_L$
[see next section] as the restriction to
$\mathbb{P}(\mathscr{B}(\euA))$ of the lift $\tilde\phi$ to
$\mathscr{B}(\euA)$ of states $\phi$ on $\euA$. If $a \in
\mathscr{B}(\euA)$, then $\tilde\phi(a)$ is the expectation value
of the measurable observable $a$ in the state $\tilde\phi$ and in
particular if $p$ is a projector in $\mathscr{B}(\euA)$, then
$\tilde\phi(p)= \phi_L(p) \in [0,1]$ yields the probability that
the proposition represented by $p$ is true in the logical state
$\phi_L$.

As it will be discussed in section 3, this setting solves the
above quoted difficulty of the $W$*-algebra approach. The scheme
can be summarized by means of the following commutative diagram

\vskip 1truecm

\begin{picture}(100,120)(-52,190)
\put(47,268){\large${\mathbb P} ({\mathscr B} (\euA))$\normalsize}
\put(130,268){\large${\mathscr B} (\euA)$\normalsize}
\put(210,268){\large$\euA$\normalsize}
\put(95,206){\large[0,1]\normalsize} \put(174,206){\large${\mathbb
C}$\normalsize}

\put(93,243){\footnotesize${\phi_L}$\normalsize}
\put(168,243){\footnotesize$\widetilde{\phi}$\normalsize}
\put(187,243){\footnotesize$\phi$\normalsize}

\put(103,268){\large$\stackrel{i}{\longrightarrow}$\normalsize}
\put(170,268){\large$\stackrel{j}{\longleftarrow}$\normalsize}
\put(135,206){\large$\stackrel{k}{\longrightarrow}$\normalsize}

\put(210,260){\vector(-3,-4){28}}
\put(75,260){\vector(3,-4){28}}
\put(150,260){\vector(3,-4){28}}
\end{picture}
\ \\

\noindent where $i,j$ and $k$ are the obvious injections. We
remark that a consequence of this proposal is that the lattice of
elementary propositions of a physical quantum system, although always
orthomodular $\sigma$-complete it is not always complete, nor
atomic, nor Hilbertian (i.e. isomorphic to all the orthogonal
subspace of a separable Hilbert space). These specific features
are encoded in the $C$*-algebra of ``continuous bounded
observables" $\euA$ of the system. More obviously, for classical
systems $\euA$ is abelian and this implies a distributive property
for the lattice of propositions.

In the rest of this paper we will make mathematically precise the
setting described above. Although the mathematical results
presented here are not original the overall scheme
and its degree of generality to the best of our knowledge are novel.

\vskip 0.2truecm \noindent
\section{Logical States}
\vskip 0.2truecm

\noindent Let $\mathscr{L}$ be the orthomodular $\sigma$-complete
lattice assumed to describe the set of elementary propositions of a physical
system. A logical state (in the sense of Mackey-Jauch-Piron
\cite{Mackey, jauch,piron}) $\phi_L$ is a $\sigma$-orthoadditive
map from ${\mathscr L}$ to [0,1]; more explicitly, if $P$ is a
proposition in ${\mathscr L}$, then $\phi_L(P^\bot)=1-\phi_L(P)$,
and if $\{P_i\}_{i \in I}$ is a countable number of propositions
pairwise orthogonal, i.e. $P_i \leq P_j^\bot$ for $i \neq j$, then
$\phi_L(\vee_i P_i)= \sum_i \phi_L(P_i)$. $\phi_L(P)$  is the
probability that the proposition $P$ is true in the state
$\phi_L$. A logical state $\phi_L$ is ``pure'' if it cannot be
written as a convex combination of other logical states i.e. if
for any two logical states $\phi_1$ and $\phi_2$ the equation
$\phi_L= \alpha \phi_1 + (1- \alpha) \phi_2, 0< \alpha <1,$
implies $\phi_L=\phi_1=\phi_2$. Pure logical states correspond to
the maximal knowledge attainable on the propositional system. A
logical state is called ``normal'' if is completely orthoadditive.

In the $W$*-algebraic approach we have the following:

\vskip 0.2truecm \noindent \textbf{Theorem 2.1.} \cite{cirelli}
\textit{Identifying as a model for ${\mathscr L}$ the lattice of
projectors of a $W$*-algebra $\euM$, the restriction to
$\mathbb{P}(\euM)$ of normal states on $\euM$ are normal logical
states; furthermore pure logical states corresponds to restriction
of pure states.}
\vskip 0.2truecm

\noindent As discussed in the introduction the proposal to
identify the logical states as restriction of normal states on
$W$*-algebras excludes the states corresponding to single points in phase space in classical systems unless the phase space is
discrete in view of the following:

\vskip 0.2truecm \noindent \textbf{Theorem 2.2.} \textit{A state
on a $W$*-algebra $\euM$ is normal iff it is an element of its
predual $\euM_*$.}

\vskip 0.2truecm \noindent Since for classical systems $\euM =
\mathscr{L}^\infty (\Omega, \omega_L)$ and $\euM_* = \mathscr{L}^1
(\Omega, \omega_L)$, this excludes the Dirac measures concentrated
on one point of $\Omega$ as they do not belong to $\mathscr{L}^1 (\Omega,
\omega_L)$.

On the other hand the requirement of normality is perfectly suited
for a standard (i.e. without, or at least with, a countable set of
superselection sectors) quantum mechanical system with finite
dynamical degrees of freedom, where we know that physical states
are ``statistical matrices", which are positive trace 1 elements of $J_1 ({\mathscr
H})$, the space of trace-class operators  on the
separable Hilbert space of physical vector states ${\mathscr H}$.
In this case, in fact, $\euM = \euB({\mathscr H})$ and $\euM_* =
J_1 ({\mathscr H})$.

The choice of a $W$* or von Neumann algebra as foundational in a
$C$* approach is also mathematically not entirely natural in NCG,
as Connes \cite{connes3} pointed out: \textit{``It is true, and at
first confusing, that any von Neumann algebra is a $C$*-algebra,
but not an interesting one because it is usually not norm
separable. For instance let $(X,\mu)$ be a diffuse probability
space (every point $p \in X$ is $\mu$-negligible), then
$\mathscr{L}^\infty (X, \mu)$ is a von Neumann algebra but it is
not norm separable and its spectrum} [see definition after Theorem
3.1 and comment after Definition 3.2] \textit{as a $C$*-algebra is
a pathological space that has little to do with the original
standard Borel space $X$''.}

The somewhat unsatisfactory situation outlined above is avoided if
we introduce the notion of Baire*-algebra.

\vskip 0.2truecm
\noindent
\section{Baire*-algebras}
\vskip 0.2truecm

To put in a proper perspective the definition of a Baire*-algebra
it is convenient to recall some basic notions of the theory of
Baire functions, whose space is the commutative version of a
Baire*-algebra.

Let $(X, \Sigma$) be a measure space, where $\Sigma$ denotes a
$\sigma$-algebra of subsets of $X$. A real or complex function is
$\Sigma$-measurable if $f^{-1} (B) \in \Sigma$ for every $B$
borelian in $\mathbb{R}$ or in $\mathbb{C}$, respectively. A class
$\mathscr{F}$ of real functions over $X$ is called monotonically
sequentially complete if every limit of a monotonic sequence of
functions of $\mathscr{F}$ belongs to $\mathscr{F}$. The class of
real $\Sigma$-measurable functions is an algebra monotonically
complete $\sigma$-stable under the lattice operations of meet and
join.

Let $X$ be a locally compact topological space. A compact set of
$X$ is of type $G_\delta$ if it is a countable intersection of
open sets of $X$. The class of $G_\delta$ compacts generates the
$\sigma$-algebra $B_X$ of the Baire sets of $X$. This is the
smallest $\sigma$-algebra from which one can reconstruct the
topology of $X$ \cite{halmos}.

A real function on $X$ is called a Baire function if it is
$B_X$-measurable; a complex function is a Baire function if both
its real and imaginary part are Baire functions. The class of real
Baire functions is the smallest class including all continuous
function in $X$ and the limit of every bounded monotone sequence of them.
The class of complex Baire functions on $X$ will be denoted by
${\mathscr B} (X)$. If $X$ is a metric space then the
$\sigma$-algebra of Baire sets coincides with the $\sigma$-algebra
of Borel sets, generated by the open sets of $X$, and the Baire
functions are Borel functions. For this reason Baire*-algebras
were called Borel*-algebras in \cite{pedersen}. To each point $p
\in X$ is associated a Dirac measure $d \mu_p$ on $\mathscr{B}(X)$
with support $\{p\}$ and mass 1.

To discuss the generalization to a non-commutative setting we need
some preliminary definitions which extend to such a setting the
basic notions involved in the constructions outlined above. A
$C$*-algebra $\euA$ is called monotonically sequentially complete
if every bounded monotone sequence of the self-adjoint part of
$\euA$, $\euA_{sa}$, possesses a limit in $\euA_{sa}.$ A state
$\phi$ over a monotonically sequentially complete $C$*-algebra
$\euA$ is called $\sigma$-normal if for every bounded monotone
sequence $\{x_n \}_{n \in {\bf N}}$ in $\euA_{sa}$ we have
$$\phi (\bigvee_n x_n) = \bigvee_n \phi (x_n).$$

\vskip 0.2truecm \noindent \textbf{Definition 3.1.}
\cite{pedersen} \textit{A $C$*-algebra ${\mathscr B}$ is called a
Baire*-algebra if it is monotonically sequentially complete ad it
admits a separating family of $\sigma$-normal states.}

\vskip 0.2truecm
\noindent Notice that, as discussed below, in the
commutative case ${\mathscr B} (X)$ is a Baire*-algebra with
separating family of $\sigma$-normal states generated by the Dirac
measures $\{d \mu_p \}_{p \in X}$.

An important result connecting Baire* and $W$*-algebras is the
following:

\vskip 0.2truecm \noindent \textbf{Theorem 3.1.} \textit{If a
Baire*-algebra has a faithful representation in a separable
Hilbert space, then it is isomorphic to a $W$*-algebra.}

\vskip 0.2truecm \noindent There is a natural ``closure" of a
$C$*-algebra to obtain a Baire*-algebra. To present this
construction we need some preliminary definitions. Given a
$C$*-algebra $\euA$, let $\hat\euA$, be its spectrum, i.e. the set
of (equivalence classes of unitarily equivalent) irreducible
representations of $\euA$. Let $\phi$ be a (representative) pure
state corresponding to a point of $\hat\euA$, and by $\pi_p$ the
corresponding representation. The atomic representation of $\euA$
is given by $\pi_a = \oplus_{\phi\in \hat\euA} \pi_\phi$ and it is
a faithful representation of $\euA$.

Then we have the following:

\vskip 0.3truecm \noindent \textbf{Definition 3.2.} (Baire*
enveloping algebra)\cite{pedersen} \textit{Given a $C$*-algebra,
$\euA$, and a subset $M \subset \euA_{sa}$, we define the monotone
sequential closure of $M$, ${\mathscr B} (M)$, as the smallest
subset of the atomic representation $\pi_a (\euA_{sa})$,
containing $\pi_a(M)$ and the limit of every monotone sequence of
elements of $\pi_a (M)$. The Baire* enveloping algebra of $\euA$,
is given by
$${\mathscr B} (\euA) \equiv  {\mathscr B}(\euA_{sa}) + i {\mathscr
B}(\euA_{sa}).$$ \noindent ${\mathscr B} (\euA)$ is a
Baire*-algebra with the family of $\sigma$-normal states given by
the unique extension of the states on $\euA$ to ${\mathscr B}
(\euA)$.}

\vskip 0.2truecm \noindent To better understand the meaning of the
Baire* enveloping algebra notice that if $\euA$ is commutative and
separable, then by the Gel'fand isomorphism (see e.g.
\cite{pedersen, thirring}), the spectrum $\hat\euA$ is a locally
compact Hausdorff space and $\euA$ is isomorphic to $C_0
(\hat\euA)$, the space of continuous function in $\hat\euA$
vanishing at infinity (if $\hat\euA$ is non-compact). Therefore
${\mathscr B} (\euA) = {\mathscr B} (\hat\euA)$, i.e. the
enveloping Baire*-algebra is exactly the algebra of complex Baire
functions on $\hat\euA$. Conversely if $\euA=C(X)$ with $X$
locally compact, $\hat\euA \simeq X$ as a topological space and
$\widehat{\mathscr B} (\euA) \simeq X$ as a Borel space. The
irreducible representations correspond to pure states given by the
normalised Dirac measures $\{d\mu_p \}_{p \in \hat\euA}.$

Notice that since ${\mathscr B} (\euA)$ has no faithful
representations on a separable Hilbert space unless $\hat\euA$ is
discrete, then in general the commutative Baire*-algebra
${\mathscr B} (\euA)$ is not a $W$*-algebra. However we have the
following result refining the previous one:

\vskip 0.2truecm \noindent \textbf{Theorem 3.2.} \cite{davies}
\textit{If $\euA$ has a faithful representation $\pi$ on a
separable Hilbert space then ${\mathscr B}(\pi(\euA)) \simeq
\pi(\euA)''$ i.e. it is isomorphic to the von Neumann algebra
generated by $\pi(\euA)$ and its $\sigma$-normal states are the
normal states of the von Neumann algebra.}

\vskip 0.2truecm \noindent For the logical interpretation, the
crucial property of Baire*-algebras is the following:

\vskip 0.2truecm \noindent \textbf{Theorem 3.3.} \textit{The set
of projectors ${\mathbb P}({\mathscr B})$ of a Baire*-algebra
${\mathscr B}$ is an orthomodular $\sigma$-complete lattice.}

\vskip 0.2truecm \noindent Furthermore, since the extensions to
${\mathscr B} (\euA)$ of the states on $\euA$ are $\sigma$-normal,
we have:

\vskip 0.2truecm \noindent \textbf{Proposition.} \textit{The
restriction of the $\sigma$-normal states of the Baire*-enveloping
algebra ${\mathscr B}(\euA)$ to  ${\mathbb P}({\mathscr B}(\euA))$
are logical states.} \vskip 0.2truecm \noindent The identification
of Baire*-algebras as the abstract setting for bounded measurable
observables is the one that makes it transparent the
interpretation of quantum mechanics as a ``theory of quantum
probability". Although there is a high amount of papers written on
this topic, it seems that a framework like the one we are
outlining here is not considered. As an example, in a quite recent
general review on the subject \cite{streater}, R. F. Streater
pointed out that: \textit{``Though the classical axioms were yet
to be written down by Kolmogorov, Heisenberg, with help of the
Copenhagen interpretation, invented a generalisation of the
concept of probability, and physicists showed that this was the
model of probability chosen by atoms and molecules."} However, the
algebraic ($W^*$-)approach envisaged therein appears less close
than ours to the standard treatment of probability on topological
measure spaces, where the Borel or Baire structure is determined
by the topology, as ${\mathscr B}(\euA)$ is determined by $\euA$.

We end this section with a

\vskip 0.2truecm \noindent \textbf{Remark.} In the definition of
enveloping Baire*-algebra we can replace the atomic representation
$\pi_a$ with the universal representation $\pi_u = \oplus_{\phi\in
S(\euA)} \pi_\phi$, where $S (\euA)$ is the set of states on
$\euA$ and the corresponding ${\mathscr B} (\euA)$ is isomorphic
to the one defined via $\pi_a$. Then ${\mathscr B} (\euA) \subset
\pi_u (\euA)''$, which is the universal enveloping von Neumann
algebra of $\euA$.

\noindent Therefore the $\sigma$-complete orthomodular lattice of
${\mathscr B} (\euA)$ describing the elementary propositions of
the system characterized by $\euA$ can be embedded in the complete
orthonormal lattice of $\pi_u (\euA)'' $; for the relevance of the
existence of the embedding from the logical point of view see
\cite{dallachiara}.

\vskip 0.3truecm

\noindent
\section{Consequences for the logic of physical systems}

\vskip 0.2truecm

\noindent Using the notions introduced in the previous section one
can make precise the scheme outlined in the Introduction.
\noindent At the foundational level one considers the algebra of
``continuous bounded observables" of the physical system,
described by a $C$*-algebra $\euA$, possibly given as the closure
of a pre-$C$*-algebra of ``smooth observables'', and the states on
$\euA$ giving the expectation values of the observables. \noindent
The algebraic realization of the lattice of elementary
propositions corresponding to yes-no experiments, concerning the
system described by $\euA$ is given by the $\sigma$-complete
orthomodular lattice of the projectors of the Baire* enveloping
algebra ${\mathscr B}( \euA)$, i.e. ${\mathbb P} ({\mathscr
B}(\euA))$. Logical states are given by the restriction to
${\mathbb P}({\mathscr B})$ of the lift to ${\mathscr B}(\euA)$ of
the algebraic states on $\euA$. Then pure logical states
describing maximal knowledge correspond to pure states on $\euA$;
notice that in general they are not pure states of ${\mathscr B}(
\euA)$.

Let us comment on some implications of the above scheme
for the logic of elementary propositions of physical systems.

\vskip 0.3truecm
\noindent \textbf{1) Systems in classical
mechanics.}

\vskip 0.2truecm

\noindent If the phase space $\Omega$ of the system is a locally
compact Hausdorff space, then $\euA = C_0 (\Omega)$ and ${\mathscr
B}(\euA) = {\mathscr B} (\Omega)$. The states on $\euA$ are the
regular Borel probability measures which have a unique extension
to ${\mathscr B} (\Omega)$. Pure states are Dirac measures
$\{d\mu_p\}_{p\in \Omega}$ with support on one point in phase space
, hence solving the problem outlined in the
Introduction.

\vskip 0.2truecm \noindent \textbf{Remark.} This solution was
first envisaged in \cite{davies,plymen} where instead of Baire*
enveloping algebras, $\Sigma$* enveloping algebras were used,
roughly speaking replacing monotone sequential closure with weak
sequential closure. In particular in the abelian case the two
concepts coincide.

\vskip 0.2truecm \noindent The lattice of propositions ${\mathbb
P} ({\mathscr B} (\Omega))$ is both atomic and distributive. As
always in the algebraic setting, there is a direct correspondence
between the abelian structure of the algebra of observables
characterising their classical nature and the distributive
property of the lattice of elementary propositions.

\vskip 0.3truecm
\noindent\textbf{2) Quantum mechanical system with
countable superselection sectors.}

\vskip 0.2truecm
\noindent \underbar{Example: quantum mechanics of
an elementary particle without spin.}

\noindent The algebra $\euA$ is the $C$*-algebra generated by the
Weyl commutation relations (1) and it is isomorphic to
$\euK({\mathscr H})$ with ${\mathscr H}$ separable infinite
dimensional; in view of Theorem 3.2, ${\mathscr B}(\euA) \simeq
\euK({\mathscr H})'' \simeq \euB({\mathscr H})$; the
$\sigma$-normal states correspond to the statistical matrices.
${\mathbb P} ({\mathscr B}(\euA))$ is atomic and Hilbertian. In
this specific example it is also irreducible, in correspondence
with the absence of superselection sectors. Notice that in the
Baire approach for classical system naturally appear the Dirac
measures excluded in the $W$* approach, whereas in quantum
mechanics are naturally excluded the singular, i.e. non-normal,
states of the above approach. By the way, our approach also
provides a natural justification for the choice made e.g. in
\cite{duvenhage} (see also \cite{clifton} for a variant) to discuss information
theory in the algebraic setting using measurable functions in
classical mechanics and bounded operators in quantum mechanics.

\vskip 0.2truecm \noindent \textbf{Remark.} The Baire approach
permits also to avoid a problematic feature appearing in the
definition of states in the temporal logic approach proposed in
\cite{primas}, where, motivated by ontological considerations
(which of course one may not agree with), a distinction is made
between ``ontic'' states and ``epistemic'' logical states. Let
${\mathscr L}$ be the orthomodular $\sigma$-complete lattice
assumed to describe the set of proposition of a physical system.
An ``ontic'' state is a lattice ortho-homomorfism $\rho$ of a
maximal orthomodular sublattice ${\mathscr T}$ of ${\mathscr L}$
into ${\mathscr B}_2$, the Boolean algebra of truth values. The
requirement on ${\mathscr T}$ to be maximal means that it does not
exist an orthomodular sublattice ${\mathscr T}^\prime$, containing
properly  ${\mathscr T}$, to which $\rho$ can be extended as
ortho-homomorphism in ${\mathscr B}_2$. This requirement
corresponds to the physical intuition of a state with ``maximal
information" and in the algebraic approach these are the pure
algebraic states.
An ``ontic'' state is called normal if  $\rho$ is a
$\sigma$-homomorphism. In this approach an ``ontic'' state refers
to ``actualized'' properties the system has (at some time). States
which refer to our knowledge are called ``epistemic''. On this
basis, if ${\mathscr L}$ is the lattice of projectors of a
$W$*-algebra $\euM$, ontic states are identified with (arbitrary,
even non normal) pure states and epistemic states with normal
states. Therefore ``ontic" states are not a subset of
``epistemic'' states. Furthermore only for normal states it has
been proved that every ontic states on ${\mathbb P}(\euM)$ has a
unique extension to a pure state of $\euM$ and every pure state on
$\euM$ defines a unique ontic state. For non normal states the
situation appear obscure, in particular for $W$*-algebras that do
not admit pure normal states! Instead in the Baire approach, i.e.
if ${\mathscr L}= {\mathbb P} ({\mathscr B}(\euA))$, one could
simply identify ``epistemic'' states with the $\sigma$-normal
states and the ``ontic'' would be those corresponding to the lift
of the pure states on $\euA$, thus a subset of the epistemic.

\vskip 0.3truecm
\noindent \textbf{3) Quantum mechanical system
with non countable superselection sectors.}

\vskip 0.2truecm
\noindent \underbar{Example: quantum mechanics of
an elementary particle without spin on a circle $S^1$.}

The algebra $\euA$ is the $C$*-algebra generated by the Weyl
commutation relations
$$e^{i n \varphi} e^{i \beta p} = e^{i \beta p} e^{i n \varphi} e^{{i \bar h\over 2}n \beta}$$

\noindent where $\varphi$ is the angle parametrizing the circle
$S^1, n \in {\mathbb Z}, \beta \in [0, 2 \pi]/\hbar$.

Inequivalent irreducible representations are labelled by an angle
$\theta \in [0, 2\pi)$ and the corresponding Hilbert space will be
denoted by ${\mathscr H}_\theta$, see e.g. \cite{thirring}. These
are the so-called $\theta$-sectors and they arise physically e.g.
in models where the particle is charged and coupled to a vector
potential whose magnetic field strength is supported in a region
in the interior of the disk bounded by circle $S^1$, in the region
forbidden for the particle motion.

A magnetic flux $\Phi$ through the disk induces a representation
of $\euA$ labelled by $\theta = \Phi$ mod $2\pi$. Hence $\euA
\simeq \oplus_\theta \euK ({\mathscr H}_\theta) \simeq C(S^1, \euK
({\mathscr H}))$ with ${\mathscr H}_\theta$ and ${\mathscr H}$
separable infinite dimensional. ${\mathscr B} (\euA) \simeq
{\mathscr B} (S^1, \euB({\mathscr H})),$ the Baire (or Borel)
functions on $S^1, \euB ({\mathscr H})$-valued.

${\mathbb P} ({\mathscr B}(\euA))$ is atomic, coincides with the
lattice of closed subspace of $\oplus_\theta {\mathscr H}$, but is
not the usual Hilbert lattice of Hilbert Quantum Logic, since
$\oplus_\theta {\mathscr H}_\theta$ is not separable, so that in
particular the lattice is not complete.

\vskip 0.3truecm
\noindent \textbf{4) Local observable algebras in
massive RQFT.}

\vskip 0.2truecm \noindent The algebraic description of (massive)
Relativistic Quantum Field Theory (RQFT) is based on the following
structure \cite{haag}: an inclusion preserving map  ${\mathscr O}
\rightarrow {\mathscr A}({\mathscr O})$ assigning to each finite
contractible open region (or alternatively open double cone)
${\mathscr O}$ in Minkowski space-time, ${\bf M}_4$, the abstract
$C$*-algebra of observables measurable in ${\mathscr O}$. The
$C$*-algebra generated by the net $\{{\mathscr A}({\mathscr
O})\}_{{\mathscr O} \subset {\bf M}_4}$ via inductive limit and
norm closure is denoted by ${\mathscr A}$ and is called the
algebra of quasi-local observables. Locality holds: if  ${\mathscr
O}_1$, ${\mathscr O}_2$ are spacelike separated, then  ${\mathscr
A}({\mathscr O}_1)$ commute with ${\mathscr A}({\mathscr O}_2)$
elementwise.

\vskip 0.2truecm \noindent \textbf{Remark.} It would be
interesting to translate the causal structure underlying the
observable net, due to a universal maximal velocity of propagation
of information, i.e. $c \neq \infty$, purely in logical terms,
like the non-distributivity of the propositional lattice in
quantum systems reflects the limitations imposed by $\hbar \neq
0$. Relevant steps in this direction can be found in
\cite{haag,mundici}.

\vskip 0.2truecm \noindent The elements of the Poincar\'e group
${\mathscr P^\uparrow_+}$ act as automorphisms on the net
preserving the local structure. Among the irreducible
representations of ${\mathscr A}$ on a separable Hilbert space in
which the Poincar\'e group is unitarily implemented, there is one,
$\pi_0$, called the vacuum representation (for simplicity assumed
unique) containing a ray, the vacuum, invariant under the unitary
representation of ${\mathscr P^\uparrow_+}$. In infinite systems,
as the one considered in RQFT, it appears in concrete examples
that physically one should not consider the set of all the
representations, but only a subset of ``physically realizable''
ones. The properties of RQFT at zero temperature and density are
discussed in terms of the net $\{\euA({\mathscr
O})=\pi_0({\mathscr A}({\mathscr O}))\}_{{\mathscr O} \subset {\bf
M}_4}$. $\euA({\mathscr O})$ can be identified as the ``space of
bounded continuous observables in the vacuum representation
measurable in ${\mathscr O}$'' .  In view of Theorem 3.2, ${\mathscr
B} (\euA({\mathscr O})) \simeq \pi_0({\mathscr A}({\mathscr
O}))''$ (and are these concrete algebras
that appear in
the constructive approach to RQFT in low dimensions \cite{glimm}); since these algebras are von Neumann algebras, ${\mathbb
P}({\mathscr B} (\euA({\mathscr O})))$ is a complete
lattice.
A deep result of RQFT with mass gap is that $\pi_0({\mathscr
A}({\mathscr O}))''$ for ${\mathscr O}$ a double cone is a type
III${}_1$ von Neumann algebra \cite{fredenhagen},
conjectured on physical grounds to be a factor
\cite{haag}. Hence the associated lattice of propositions is
non-atomic, the projectors having Murray-von Neumann dimensions
only {0, $\infty$}. In the Baire approach the $\sigma$-normal
states are the normal states of $\pi_0({\mathscr A}({\mathscr
O}))''$, however a factor III${}_1$ does not possess pure normal
states. Nevertheless in our approach pure logical states
corresponding to maximal knowledge on the proposition lattice of
the local system are naturally defined, as they are obtained from
lifts of states on $\pi_0({\mathscr A}({\mathscr O}))$, which
being a $C$*-algebra with unity has a separating family of pure states.

\vskip 0.3truecm
\noindent
\section{Conclusions}
\vskip 0.2truecm

\noindent Summarizing, in this paper we propose that the lattice
of elementary propositions of physical systems is completely
encoded in the $C$*-algebra $\euA$ of ``continuous bounded
functions or observables'' on a generally ``non-commutative phase
space $X$'' in the sense of Non Commutative Geometry. The
propositional lattice can be represented as the $\sigma$-complete
orthomodular lattice of projectors of the space of ``(Baire)
measurable bounded observables on $X$'', which can be obtained as
a suitable closure, via the Baire envelope, of $\euA$. Hence the
propositional logic depends on the physical system, but it
captures only a very ``coarse grained'' structure of it. For
example it is able to identify the classical or quantum nature of
the system and it is sensible to the related ``completeness'' or
``incompleteness'' through the verification of the validity of the
Lindenbaum property \cite{giuntini} in the corresponding logic.
But it is also able to distinguish more refined features of
quantum systems e.g. the presence of a countable from a
non-countable set of superselection sectors or the ``dimension''
in the sense of Murray-von Neumann of the sectors.

\bigskip

\paragraph{Acknowledgements.}
We gratefully acknowledge R. Nobili for many illuminating
discussions and for a critical reading of a preliminary version of
the manuscript. This work is supported in part by the European Community's Human Potential Programme under contract HPRN-CT-2000-00131 Quantum Spacetime.

\vskip 0.2truecm

\vskip 0.3truecm
\noindent
\appendix
\section{Appendix}
\vskip 0.2truecm

\noindent \textbf{$\mathbf{C}$*-algebra.} A $C$*-algebra $\euA$ is
an algebra over $\mathbb{C}$, with an involution
* and a norm $|| \cdot||$; in this norm $\euA$ is
complete, i.e. Banach, and  $\forall a,b \in \euA$ $||ab|| \leq
||a|| ||b||;||a^*||=||a||$; the key property linking the algebraic
and the topological structure holds: $||a^*a||=||a||^2$ and if the
unity ${\bf 1} \in \euA$ then  $\euA$ is called unital and $ ||{\bf 1}||=1$. Every $C^*$ algebra without unity  $\euA$ can be canonically embedded in a unital $C^*$ algebra $\tilde\euA$ as an ideal satisfying $\tilde\euA/\euA \simeq {\mathbb
C}$; in the following if $\euA$ is not unital ${\bf 1}$ is referring to $\tilde\euA$.  An element $a \in
\euA$ is called {\it self-adjoint} or hermitian iff $a^*=a$; {\it
projector} iff $a^2=a=a^*$; unitary iff $aa^*=a^*a={\bf 1}$;
positive iff there exists $b \in \euA$ such that $a=b^*b$; an
element $b \in \euA$ is called the inverse of $a$ iff $ab=ba={\bf
1}$ and then denoted by $a^{-1}$. The {\it spectrum} of  $a \in
\euA$ is the set $Sp(a)=\mathbb{C} \, \backslash  \{z \in
\mathbb{C} , (z-a)^{-1} \in \euA \} $. The norm of a $C^*$-algebra
can be uniquely algebraically defined as $||a||=$sup$\{|z|, z \in
Sp(a^*a)\}^{1/2}$.

In a $C$* approach the bounded physical observable quantities of a
physical system are described by the self-adjoint elements of a
$C$*-algebra and the possible results of a measurement on the
physical observable described by $a$ are given by the spectrum of
$a$.

\vskip 0.2truecm \noindent \textbf{State.} An algebraic state
(here simply called state) on $\euA$ is a positive linear
functional $\phi$ on $\euA$, normalized by $\phi({\bf 1})=1$.
Convex combinations of states are states. States that cannot be
written as convex combination of other states are called {\it
pure}. A family $F$ of states is called {\it separating} if
$\phi(a)=0$ for all $\phi \in F$ implies $a=0$ for all positive $a
\in  \euA$. Every unital $C$*-algebra has a separating family
of pure states.

In a $C$* approach the (algebraic) states describe the ``states of
knowledge" of the observable quantities and pure states correspond
to maximal knowledge. The expectation value of the measures
performed on the physical observable described by $a$ in the state
of knowledge described by $\phi$ is given by $\phi(a)$.

\vskip 0.2truecm \noindent \textbf{Representation.} Let $\euA$ be
a $C$*-algebra, ${\mathscr H}$ a Hilbert space and $\euB({\mathscr
H})$ the $C$*-algebra of bounded operators on ${\mathscr H}$. A
{\it representation} $\pi$ on ${\mathscr H}$ is a homomorphism of
$\euA$ into $\euB({\mathscr H})$ preserving the involution. If
${\mathscr S}$ is a $\ast$-subalgebra of $\euB({\mathscr H})$,
${\mathscr S}'$ denotes its commutant, i.e. the set of elements of
${\mathscr B}({\mathscr H})$ commuting with all the elements of
${\mathscr S}$. A representation $\pi$ is called {\it faithful}
iff $\pi(a)=0$ implies $a=0$; {\it irreducible} if the commutant
$\pi(\euA)'$ contains only multiples of the unity; two
representations $\pi_1$ on  ${\mathscr H}_1$ and $\pi_2$ on
${\mathscr H}_2$ are called {\it unitarily equivalent} if there
exists an isometry $u$ of ${\mathscr H}_1$ onto ${\mathscr H}_2$
such that $u \pi_1(a) u^* =\pi_2(a), \forall a \in  \euA$.

\vskip 0.2truecm \noindent \textbf{von Neumann algebra.} A weakly
closed $\ast$-subalgebra $\euM$ of $\euB ({\mathscr H})$ is called
a von Neumann algebra. The von Neumann double commutant theorem
states that $\euM= \euM''$; more generally if ${\mathscr S}$ is a
$*$-subalgebra of ${\mathscr B}({\mathscr H})$, then ${\mathscr
S}''$ is called the {\it von Neumann algebra generated by}
${\mathscr S}$. A von Neumann algebra $\euM$ is called a {\it
factor} iff the centre $\euM \cap \euM'$ contains only multiples
of the unity.

\vskip 0.2truecm \noindent \textbf{$\mathbf{W}$*-algebra.} A
$W$*-algebra $\euM$ is a $C$*-algebra which in addition is the
dual of a Banach space, called its {\it predual} and denoted by
$\euM_*$. The dual space  $\euM^*$ of linear functionals on $\euM$
is larger then the predual, hence the set of states on $\euM$ have
a distinguished subset contained in the predual; these are the
{\it normal states}; they are completely additive on projectors of
$\euM$. Every $W$*-algebra $\euM$ admits a faithful representation
as a von Neumann algebra in some Hilbert space ${\mathscr H}$;
${\mathscr H}$ can be taken separable iff the predual $\euM_*$ is
norm separable.

\vskip 0.2truecm \noindent \textbf{Murray- von Neumann dimension.}
 Two projectors $p_1$ and $p_2$ in a factor $\euM$,
projecting onto subspaces ${\mathscr H}_1$ and ${\mathscr H}_2$ of
${\mathscr H}$ are said equivalent iff there exists a partial
isometry $V \in \euM$ from  ${\mathscr H}_1$ to ${\mathscr H}_2$,
i.e $p_1=V^*V, p_2=VV^*$ and then we write $p_1 \sim p_2$.
 One can order the equivalence class of projectors by
setting $p_1 < p_2$ iff $p_1 \nsim p_2$ and there exists a proper
subspace of ${\mathscr H}_1$ whose associated projector is
equivalent to $p_2$. A projector $p_1$ is called finite iff $p
\leq p_1$ and $p \sim p_1$ implies $p = p_1$. There exists a
positive function on the equivalence classes of projectors, the
{\it Murray-von Neumann dimension} $d$, satisfying $d(0)=0$,
$d(p_1)=d(p_2)$ iff $p_1 \sim p_2$, $d(p_1) < d(p_2)$ iff  $p_1 <
p_2$ and, if $p_1p_2=0$, $d(p_1+p_2)=d(p_1)+d(p_2)$. For factors
with separable predual the following alternatives exists: a factor
is of type I if it contains atoms, i.e. minimal nonzero
projectors, whose von Murray-von Neumann dimension is 1 and the
range of $d$ is a subset of ${\mathbb N}$, in particular it is
called of type I$_n$ if $n$ is the maximal value in the range of
$d$; of type II if it is atom-free and it contains some nonzero
finite projector; of type III if it does not contain any nonzero
finite projector and then $d$ takes only the values $0$ and
$\infty$.

\end{document}